# Advanced configuration of gravitational-wave interferometer on the base of "sensitive mode" in "white-light cavity"

G. G. Karapetyan

*Yerevan Physics Institute, Armenia 375036, E-mail:gkarap@crdlx5.yerphi.am*

**Abstract**
 A novel conception of "sensitive mode" (SM) is proposed to apply in gravitational-wave advanced interferometer configuration (AIC). The SM is resonant oscillation of electromagnetic field in "white-light cavity" (WLC), where the resonance line is broadened without decreasing cavity quality. The resonant field frequency of the SM is greatly susceptible to the change of cavity length, and the SM is established in a cavity with time constant smaller than a conventional mode. Due to these advantages the sensitivity and bandwidth of AIC can be increased

There are currently several projects constructing GW detector that relay on ground-based laser interferometers: TAMA, GEO, VIRGO, and LIGO [1]. In the VIRGO and the LIGO they expect to measure a small phase shift between two light beams, traversing two orthogonal arms of interferometer. The length of the arm is arranged such that path lengths in two arms differ by a half wavelength, so the two beams interfere in output destructively, producing a dark fringe. However, a passing GW would stretch and squeeze the arms, causing a small phase shift, which is converted into an observable light-power change at interferometer output. Different noise sources deteriorate output signal-to-noise ratio [2]. To increase phase shift in the LIGO two Fabry-Perot cavities are used in the arms, where light is reflected back and forth several times increasing effective path length. Similar effect will be achieved in signal recycling configuration [3] of the GEO, where an additional cavity is formed by a partly transparent mirror, placed in interferometer output. Other long time development topics to increase the performance of the GW interferometer are using of cryogenic bar detectors, squeezed light, QND technique and white-light cavity (WLC) [4]. The concept of the WLC has been proposed to increase the bandwidth of interferometer [5]. For this purpose a transparent medium with anomalous dispersion was suggested to use in the cavity. At some value of dispersion slope the resonance line considerably broadens (without decreasing of cavity quality), providing by this the increasing of interferometer bandwidth. In a recent experiment [5] a medium with anomalous dispersion and low losses was prepared, proving that a transparent medium with anomalous dispersion can be realized.
In this present paper we show that except of broadened line the oscillation in WLC is extremely sensitive to the change of cavity length, and define such oscillation as the SM. We derive analytical formulas describing the parameters of the SM and analyze the AIC based on the SM, as well as propose a new way to establish the SM by using a medium with normal dispersion.
Consider a Fabry-Perot cavity with an input mirror intensity reflection coefficient $r^2$ and losses p. Instead of a rear mirror a small cavity with input mirror intensity reflection coefficient $r_1^2$ and losses $p_1$ is used. The rear mirror of the small cavity is perfectly reflecting. The lengths of the large and the small cavities are correspondingly L and b (see Fig. 1). The small cavity is filled with a gas with refractive index $n(\omega)$, and absorption coefficient $\alpha(\omega)$. The reflective function of the cavity for monochromatic waves, calculated by using standard techniques [6] is

$$R(\omega) = ir + i\frac{1 - r^2 - p}{r + (i/R_1(\omega))\exp(-2iL\omega/c)}, \quad (1)$$

where c is the speed of light in vacuum, and $R_1(\omega)$ is the reflective function of the small cavity, which can be calculated analogously

$$R_1(\omega) = ir_1 + i\frac{1 - r_1^2 - p_1}{r_1 + \exp[2b\omega(\alpha(\omega) - in(\omega))/c]}. \quad (2)$$

The lengths of cavities are arranged to provide resonance conditions $\exp(-2iL\omega_0/c) = -1$, and $\exp(-2ib\omega_0 n(\omega_0)/c) = -1$, as well as the reflectivity of the small cavity is arranged to be about unity (see below). When both cavities resonate at angular frequency $\omega_0$, the reflective function (1) is:

$$R(\omega) = ir + i\rho(\omega), \quad (3)$$

$$\rho(\omega) = \frac{1 - r^2 - p}{r - (1/|R_1(\omega)|)\exp[-i\psi(\omega, L)]}, \quad (4)$$

$$\psi(\omega, L) = 2L\omega/c + \varphi(\omega) - \pi/2, \quad (5)$$

where $\varphi(\omega) = \arg R_1(\omega)$.

Here $|\rho(\omega)|$ is a narrow resonance function with central angular frequency $\omega_0$. As it is seen from (4) $\omega_0$ is a root of the equation $\psi(\omega, L) = 2\pi k$ (k is an integer). When the cavity length changes by $\Delta L$ the resonance angular frequency changes by $\Delta\omega$, and its new value $\omega_0 + \Delta\omega$ must satisfy the equation

$$\psi(\omega_0 + \Delta\omega, L + \Delta L) = 2\pi k. \quad (6)$$

To approximately solve the (6) an reveal the relationship between $\Delta\omega$ and $\Delta L$, it is convenient to expand $\varphi(\omega_0 + \Delta\omega)$ in series of Taylor around $\omega_0$. Since $\varphi(\omega)$ is an odd function of $(\omega - \omega_0)$, its second derivative in $\omega_0$ is zero, and (6) becomes

$$\frac{\varphi'''}{6}\Delta\omega^3 + [2(L + \Delta L)/c + \varphi']\Delta\omega + 2\Delta L\omega_0/c = 0, \quad (7)$$

where $\varphi'$ is the first derivative of the phase (phase slope), $\varphi'''$ is the third derivative of the phase respect to $\omega$ calculated in $\omega_0$.

In the case of a cavity with a conventional rear mirror having constant phase ($\varphi' = \varphi''' = 0$) the well known expression of frequency shift follows from (7)

$$\Delta\omega = -\omega_0 \Delta L/L. \quad (8)$$

However in the case when a rear mirror (or a small cavity) has a frequency dependent phase with the negative slope complying with the condition

$$\varphi' = -2L/c, \quad (9)$$

the solution of (7) essentially differs from (8), being with relative error $\sim(\Delta L/L)^{1/3}$

$$\Delta\omega = -(12\omega_0\Delta L/c\varphi''')^{1/3}. \quad (10)$$

To estimate (10) note that negative phase slope is rather an unusual property, which takes place in a narrow frequency interval, where the phase changes on $\sim \pi$ (see Fig. 2). Since we require (9) it means that this angular frequency interval is $d\omega \sim \pi c/2L$. Therefore $\varphi''' \sim \pi/d\omega^3 \sim (L/c)^3$, and hence from (10) we obtain the approximate formula

$$\Delta\omega \sim -\omega_0 (\Delta L / L)^{1/3} (\lambda_0 / L)^{2/3}, \qquad (11)$$

where $\lambda_0 = 2\pi c/\omega_0$ is the wavelength.

Note that in [7], to estimate $\varphi''$ (if $\varphi$ is not an odd function) we used the crude formula $\varphi'' \sim L/c\omega$ obtaining an overestimated value of frequency shift $\Delta f \sim f(\Delta L/L)^{1/2} \sim 10$ kHz for $\Delta L/L = 10^{-21}$. By using present the more correct approximation of the frequency interval where the negative phase slope takes place, one can obtain $\varphi'' \sim \pi/d\omega^2 \sim (L/c)^2$, and better estimation of the frequency shift in that case: $\Delta f \sim f(\Delta L/L)^{1/2}(\lambda_0/L)^{1/2} \sim 0.1$ Hz.

Formula (11) provides a good estimation for analyzing $\Delta f$. It shows that the frequency shift decreases, when cavity length increases. Nevertheless, due to the cubic root relationship, the value of (11) even for $L = 4$ km still considerably surpasses conventional frequency shift (8). For example, if $L = 4$ km, $\lambda_0 = 1064$ nm, and $\Delta L/L = 10^{-21}$, formula (11) gives $\Delta\omega/2\pi \sim 10$ Hz, meanwhile a conventional frequency shift (8) is only about $10^{-7}$ Hz.

Thus under condition (9) a special oscillation is established in cavity, being extremely susceptible to the change of cavity length. When this condition is violated and the second term in (7) becomes equal to the third one in the order of smallness, the frequency shift sharply decreases to the conventional value (8), which means that the SM is transformed to conventional mode. Note that condition (9) was proposed (in other definitions) for the first time in [5] to increase the bandwidth of cavity without decreasing its quality. Such novel cavity was defined as the WLC to emphasize that it resonates in wide band of frequencies. Here we showed another important consequence of (9) missed in [5]: under condition (9) the oscillation becomes extremely susceptible to the change of cavity length. Because of that, we define such oscillations as the SM and believe that the SM can find different applications, particularly in interferometric devices. Thus, the SM is actually the resonant oscillations in the WLC. Nevertheless, introduction of a new term – the SM seems to be relevant, because the SM can be established not only in optical resonators, but for example in microwave [8] or acoustic resonators. On the other hand, the terminology WLC does not point out on the high susceptibility of oscillations in the WLC.

Negative phase slope, being the key condition for establishing the SM (or for building the WLC) can be realized by using a transparent medium (gas) having anomalous dispersion [5, 8] (it should be noted that such gas has been investigated in the past [9] as a medium supporting superluminal propagation of light pulses). Here to obtain negative slope we propose to use in the small cavity a gas, which is pumped to obtain the population inversion. Refractive index and amplification coefficient of such gas with a spectral line centered on the $\omega_0$ can be written as a conventional (absorbing) gas using a simple oscillator model. Assuming that all the atoms are pumped to a given energy level we will have

$$n(\omega) + i\alpha(\omega) = 1 + \frac{2\pi A}{\omega - \omega_0 + 2\pi i\gamma} \; . \qquad (12)$$

Here $A = Ne^2/m\omega_0$, $N$ is density of the gas, $e$ and $m$ are the charge and mass of electron and $\gamma$ is a damping constant for which we will use an arbitrary value 0.5 MHz.

Although in the vicinity of $\omega_0$ such gas has normal dispersion and a positive phase slope, nevertheless by appropriately arranging the small cavity parameters one can achieve required negative phase slope satisfying (9). For this purpose, we will use $r_1 =$ 0.4, $p_1$ = 0.8384243, $Ab/\lambda_0$ = 36720.14 Hz. The last condition can be arranged both by tuning the length of the small cavity or the density of the gas. For example, if b = 10 cm and $\lambda_0$ = 1064 nm, it must be A = 0.3907023 Hz. Then, calculating in (2) the reflective function of the small cavity we have $|R_1(\omega_0)|$ = 1.00000175, and $\varphi'$ = $-$26.67·10$^{-6}$ sec, satisfying eq. (9) for L = 4 km. Thus, the SM is established in the large cavity or, which is equivalent, the large cavity becomes the WLC, providing greatly increased value of frequency shift (10). Calculating from (4), and substituting in (10) the value $\varphi'''$ =1.63·10$^{-13}$ sec$^3$ we obtain the exact formula for the shift $\Delta f$ of the SM resonance frequency

$$\Delta f[Hz] = \Delta\omega/2\pi = 1.95 \cdot 10^7 (\Delta L/L)^{1/3}. \qquad (13)$$

For $\Delta L/L$ = 10$^{-21}$ this formula gives $\Delta f$ = 1.95 Hz, unlike to above mentioned approximated value 10 Hz, following from (11). Note, that (10), (11), (13) are still valid if $\Delta\omega$ is inside the frequency interval $d\omega \sim \pi c/2L$, where the phase has a negative slope. By substituting this value in (11), one can found that the maximum shift of the cavity length $\Delta L_{max}$ still allowing the existence of the SM is $\Delta L_{max} \sim 0.1\lambda_0$. Further shift of the cavity length the cubic root dependence (11) transforms to linear one (8), which means that the SM is transformed to conventional mode, or the WLC becomes a conventional cavity. This conclusion follows from the exact numerical solution of eq. (6), which gives the SM resonance frequency shift versus the change of the cavity length (see Fig. 3).

On the base of high susceptibility of SM frequency to the change of cavity length an AIC scheme can be proposed where the shift of SM frequency is converted to appropriately increased phase shift. In this AIC the external laser is removed, and the lasing of the SM originates in the large cavity, when the gas in the small cavity is appropriately pumped. By using an additional transparent mirror (is shown in Fig. 1 by the dotted line) a small portion of the light can be extracted from the large cavity in two opposite directions. The phase difference $\Phi$ of these two beams is $\Phi = 2\omega_0 d/c$, where d is the distance between the additional mirror and the input mirror of the large cavity. When cavity length will change on $\Delta L$ due to the GW, the frequency of the SM will also change according to (10), and $\Phi$ will change by $\Delta\Phi = 2\omega_0\Delta d/c + 2\Delta\omega d/c \approx 2\Delta\omega d/c$. For the parameters used in (13) and d ≈ 1 km we obtain $\Delta\Phi \approx 10^3(\Delta L/L)^{1/3}$, which gives a strongly increased value of the phase shift $\Delta\Phi \approx 10^{-4}$ rad for $\Delta L/L$ = 10$^{-21}$, and $\Delta\Phi \approx 10^{-5}$ rad for $\Delta L/L$ = 10$^{-24}$. Proposed AIC is suitable as well for the interferometers having one arm.

Another feature of the SM and WLC is the broadened resonance line, which enhances interferometer bandwidth. Let us consider this feature in more details using the above obtained parameters of small cavity, which transforms the large cavity into the WLC. Numerical calculations of (4) give the resonance function $|\rho(\omega)|$, shown in Fig. 4 (solid lines). The dotted lines correspond to the resonance function of the conventional cavity, which is obtained from (4) by substituting $|R_1(\omega)|$=1, $\varphi(\omega) = \pi/2$. It is seen that the WLC has wider resonance line and consequently shorter storage time than a conventional cavity, because cavity bandwidth and storage time are inversely proportional quantities (it follows from the general properties of Fourier transform of the cavity resonance function and its temporal function, describing the processes of loading or decaying).

It is useful to reveal the relation between the bandwidth and the finesse of the WLC. By changing r in (4), one can numerically derive the values of the finesse and appropriate bandwidths for both conventional cavity and the WLC. To estimate analytically this relation, let us expand the exponent in (4) in Taylor series with three terms around $\omega_0$. Then the width $2\delta\omega$ of the resonance function at 0.71 amplitude is determined by the equation

$$(r\beta - 1 + \delta\psi^2/2)^2 + \delta\psi^2 = 2(r\beta_0 - 1)^2, \qquad (14)$$

where $\beta_0 = |R_1(\omega_0)|$, $\beta = |R_1(\omega_0+\delta\omega)|$,.

For a conventional cavity ($\delta\psi = L\delta\omega/c$) the known expressions of the frequency bandwidth $2\delta f = 2\delta\omega/2\pi$ and the storage time $\tau$ follow from (14), by substituting $\beta \approx \beta_0 = 1$

$$2\delta f = c/2LF, \quad \tau = 1/2\pi\delta f = 2LF/\pi c, \qquad (15)$$

where $F = \pi(r\beta_0)^{1/2}/(1 - r\beta_0)$ is the cavity finesse.

However, for the WLC one should take $\delta\psi = \varphi'''\delta\omega^3/6$ and utilize the function $|R_1(\omega)|$. Then the dependence $\beta = \beta_0(1 - 810\delta\omega^2)$ is numerically derived from (2), and after appropriate transformations of (14) the following simple expression for the WLC bandwidth is obtained

$$2\delta f_{WLC}[\text{kHz}] = 11.37/F^{1/2} . \qquad (16)$$

As it was mentioned above the WLC storage time is inversely proportional to its bandwidth. However, the appropriate exact formula can not be derived within the present monochromatic waves consideration. Therefore we have the following asymptotic formula for the WLC storage time

$$\tau_{WLC}[\text{ms}] \sim 1/2\pi\delta f_{WLC} \sim 0.03F^{1/2} . \qquad (17)$$

Corresponding to (15) - (17) curves are plotted in Fig. 5 by the solid and the dotted lines, revealing that the WLC bandwidth is inversely proportional, and the storage time is directly proportional to the square root of cavity finesse. Because of that it is beneficially to employ the WLC with high finesse, to increase the GW induced output phase shift; by this the storage time increases not very much.

In conclusion, we revealed that the frequency of resonant oscillation in the WLC is extremely susceptible to the change of cavity length, and introduced for such oscillation a new term – the SM. We

- obtained asymptotic formulas for the SM main features. These formulas show that the shift of the SM resonance frequency is proportional to the cubic root upon the relative shift of cavity length, so it greatly surpasses that of a conventional mode;
- revealed that the WLC bandwidth is inversely proportional, and the storage time is directly proportional to the square root upon cavity finesse, which enhances considerably the interferometer bandwidth;
- proposed the AIC scheme with significantly increased sensitivity, where the SM frequency shift is converted to a strongly increased phase shift;
- proposed a new approach to build the WLC by using an active medium with normal dispersion.

Along with applying in GW interferometer the SM can find applications in other interferometric devices.

**Acknowledgements**
The author thanks D. Shoemaker for helpful discussions and comments on the paper. This work was supported by the grant INTAS No 97-30748.

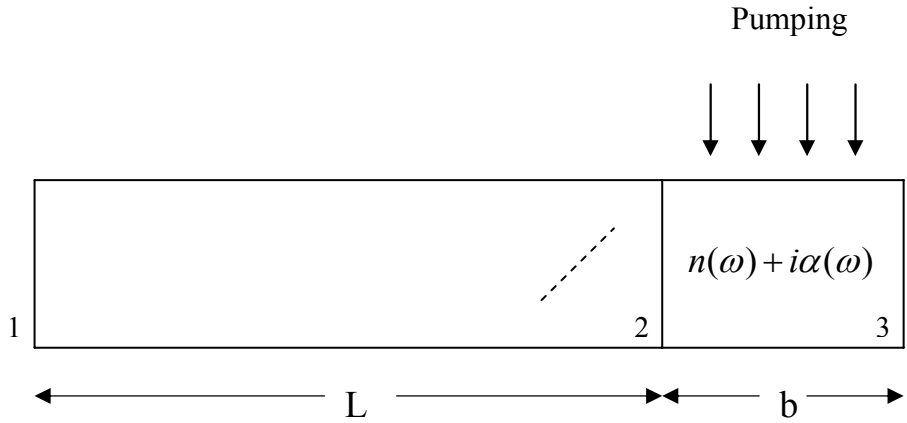

Fig. 1 Sketch of the cavity configuration. The dotted line is an additional mirror, which can be used for extracting a small portion of the light in two opposite directions. Reflective functions of the input mirror of the large cavity (1), the input mirror of the small cavity (2), and rear mirror are correspondingly ir, ir$_1$ and i.

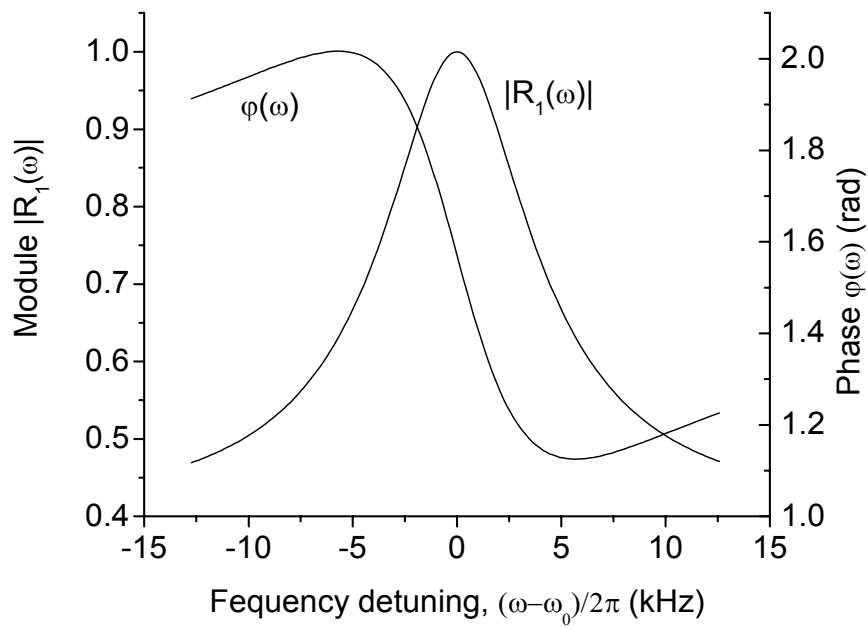

Fig. 2 Resonance function of the small cavity. $r_1 = 0.4$, $p_1 = 0.8384243$, $\gamma = 0.5$ MHz, $Ab/\lambda_0 = 36720.14$ Hz.

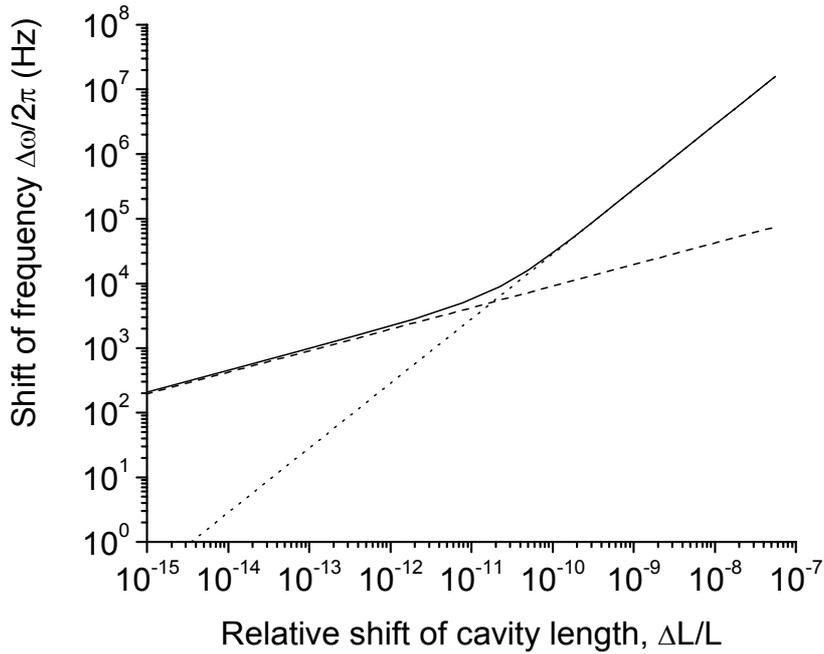

Fig. 3 Shift of the resonance frequency of the SM versus 4-km cavity length variation (solid line) derived from the exact numerical solution of eq. (6). The dotted line corresponds to linear dependence (8) of conventional mode, and the dashed line represents the asymptotic cubic root dependence (13).

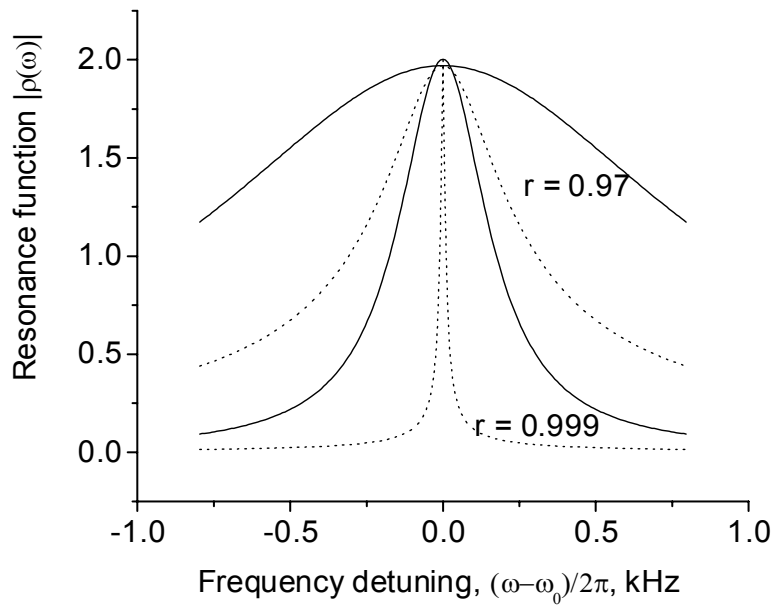

Fig. 4 Resonance functions of 4-km WLC (the solid lines) and conventional cavity (the dotted lines). For r = 0.97 (which corresponds to cavity finesse F = 103) the conventional cavity has line width at 0.71 amplitude 0.36 kHz, and the WLC has line width 1.24 kHz. For r = 0.999 (finesse is 3150) the line widths of the conventional cavity and the WLC are correspondingly 12 Hz, and 224 Hz.

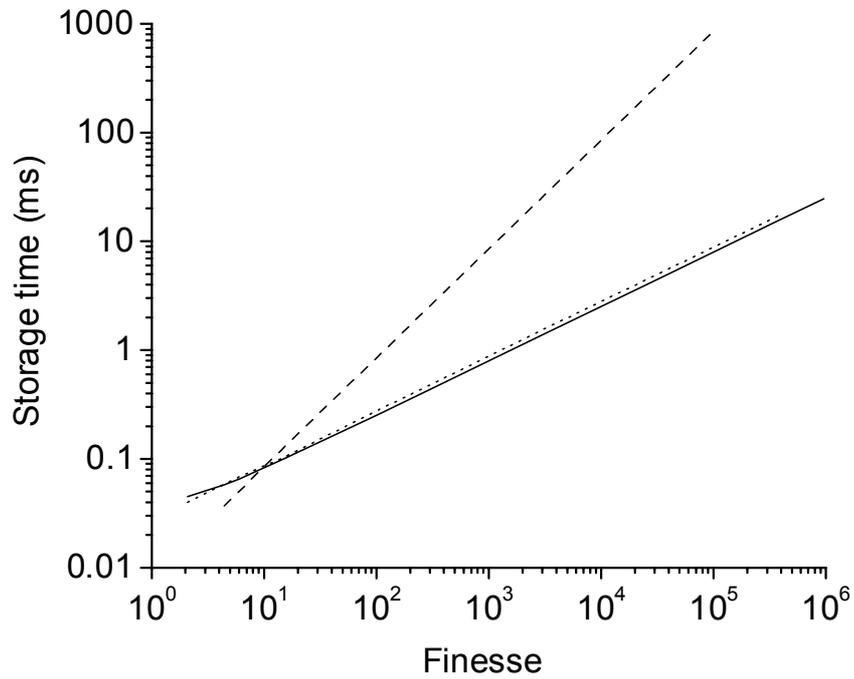

Fig. 5 Storage times of 4-km conventional cavity (the dashed line) and the WLC (the solid line), derived from numerical analysis of the resonance function (4). The dotted line corresponds to approximate formula (17). Storage times of conventional cavity and the WLC are correspondingly 0.87 ms, and ~ 0.3 ms for F = 103 (r = 0.97), and 26.67 ms and ~ 1.4 ms for F = 3150 (r = 0.999).